\documentclass[12pt]{article}

\usepackage{epsf}
\usepackage{amssymb}

\def\be{\begin{equation}}
\def\ee{\end{equation}}
\def\bea{\begin{eqnarray}}
\def\eea{\end{eqnarray}}
\def\cN{{\cal N}}
\def\dfrac#1#2{{\displaystyle\frac{#1}{#2}}}

\def\Tr{{\rm Tr}}
\def\nnabla{\nabla\!\!\!\!\nabla}
\def\ssquare{\square\!\!\!\!\square}
\def\theequation{\arabic{section}.\arabic{equation}}

\makeatletter
\long\def\@makecaption#1#2{{%
  \small
  \vskip\abovecaptionskip
  \sbox\@tempboxa{#1. #2}%
  \ifdim \wd\@tempboxa >\hsize
    #1. #2\par
  \else
    \global \@minipagefalse
    \hb@xt@\hsize{\hfil\box\@tempboxa\hfil}%
  \fi
  \vskip\belowcaptionskip}}
\makeatother

\begin{document}
\begin{titlepage}
\begin{flushright}
ITP-UH-17/06\\
\tt hep-th/0608048
\end{flushright}
\vspace{5mm}
\begin{center}
{\Large\bf
%Low-energy effective action for the
% non-anticommutative charged hypermultiplet model
Vector-multiplet effective action in the\\\vspace{0.2cm}
non-anticommutative charged hypermultiplet model}
 \\[1cm]
 {\bf
 I.L. Buchbinder $^+$\footnote{joseph@tspu.edu.ru},
 O. Lechtenfeld $^\ddag$\footnote{lechtenf@itp.uni-hannover.de},
 I.B. Samsonov $^{*}$\footnote{samsonov@mph.phtd.tpu.edu.ru}}\\[3mm]
 {\it $^+$ Department of Theoretical Physics, Tomsk State Pedagogical
 University,\\ Tomsk 634041, Russia\\[2mm]
 $^\ddag$ Institut f\" ur Theoretische Physik, Leibniz Universit\" at
 Hannover,\\ Appelstra\ss e 2, D-30167 Hannover, Germany\\[2mm]
 $^*$ Laboratory of Mathematical Physics, Tomsk Polytechnic
 University,\\ 30 Lenin Ave, Tomsk 634050, Russia
 }
 \\[0.8cm]
\bf Abstract
\end{center}
We investigate the quantum aspects of a charged hypermultiplet in
deformed $\cN{=}(1,1)$ superspace with singlet non-anticommutative
deformation of supersymmetry. This model is a ``star'' modification
of the hypermultiplet interacting with a background Abelian vector
superfield. We prove that this model is renormalizable in the sense
that the divergent part of the effective action is proportional to
the $\cN{=}(1,0)$ non-anticommutative super Yang-Mills action. We
also calculate the finite part of the low-energy effective action
depending on the vector multiplet,
which corresponds to the (anti)holomorphic potential. The
holomorphic piece is just deformed to the star-generalization of the
standard holomorphic potential, while the antiholomorphic piece is
not. We also reveal the component structure and find the deformation
of the mass and the central charge.
\end{titlepage}
\setcounter{footnote}{0}

\section{Introduction and Summary}
In this paper we study the quantum aspects of
$\cN{=}(1,0)$ non-an\-ti\-com\-mu\-ta\-ti\-ve theories with singlet
deformation of supersymmetry. To provide the motivation of this work let us
briefly summarize the most important achievements and problems concerning
non-anticommutative theories with $\cN{=}(1/2,0)$ and $\cN{=}(1,0)$
supersymmetry.

The interest in $\cN{=}(1/2,0)$ non-anticommutative
deformations of supersymmetry originated with the papers
\cite{Strings,Strings1,Strings2},
where these deformations were derived from superstring theory
on a constant graviphoton background. As a rule, such deformations break
supersymmetry only in the chiral sector of superspace, which is possible
in Euclidean superspace. The key feature of non-anticommutative deformations
on the quantum level is the preservation of renormalizability,
which was established for $\cN{=}(1/2,0)$ Wess-Zumino \cite{WZ} and
super Yang-Mills (SYM) \cite{SYM12,Jack,Penati} models.
This result is very non-trivial since the
non-anticommutative deformations involve a parameter
with negative mass dimension which plays the role of a new coupling constant.
Since such theories appear to be renormalizable, their
quantum dynamics should be explored. Indeed, the low-energy effective action
of such models was considered in \cite{B}, where the corrections due to
the non-anticommutative deformation were calculated.
These results provide a promising new method
of partial supersymmetry breaking which preserves renormalizability.
New remarkable quantum features suggest interesting physical applications
(see, e.g., \cite{Strings1} for modifications of the glueball superpotential
and the expectation value of the glueball field in $\cN{=}(1/2,0)$
supersymmetric field theories).

These surprises of $\cN{=}(1/2,0)$ theories motivated analogous investigations
of deformed {\sl extended\/} supersymmetric theories. In this case
we distinguish several types of non-anticommutative deformations.
The simplest one depends on a single scalar parameter $I$ which appears in
the anticommutator of the chiral $\cN{=(}1,1)$ Grassmann coordinates,
\be
\{\theta^\alpha_i,\theta^\beta_j \}_\star=2 I\varepsilon^{\alpha\beta}
\varepsilon_{ij}.
\label{eq1}
\ee
Such a deformation was introduced in \cite{singlet} and was named a
chiral singlet deformation; the corresponding field theories are
referred to as $\cN{=}(1,0)$ non-anticommutative.
Although more general types of deformations of extended supersymmetry
have been considered in \cite{CILQ,Araki1}, here we shall restrict ourselves
to the singlet deformation (\ref{eq1}), since this case is most elaborated
now on the classical level \cite{FILSZ}-\cite{Araki} and its stringy origins
have been established \cite{FILSZ}.

The quantum aspects of $\cN{=}(1,0)$ non-anticommutative
theories are more involved. In \cite{FILSZ,ILZ,IZ} it was shown that the
$\cN{=}(1,0)$ SYM and hypermultiplet models acquire a number of new classical
interaction terms even in the Abelian case. In principle, such terms can
spoil renormalizability. In our recent paper \cite{my} we addressed
this problem in two such $\cN{=}(1,0)$ theories, namely
the Abelian SYM model and the {\sl neutral\/} hypermultiplet interacting
with an Abelian gauge superfield. By computing all divergent contributions
to the effective action we proved that both these models remain renormalizable.
Note that these theories are deformations of free ones, and thus
all interaction terms vanish in the undeformed limit $I\to0$.
A physically more important example is the {\sl charged\/} hypermultiplet
model, which -- already prior to the deformation -- features the interaction
of a hypermultiplet with a background Abelian vector superfield.
In particular, the low-energy effective action of this theory is goverend by
the so-called holomorphic potential, which plays a significant role in
the Seiberg-Witten theory \cite{Seiberg}.
Therefore, in the present work we study the low-energy effective action and
renormalizability for the $\cN{=}(1,0)$ non-anticommutative charged
hypermultiplet model.

Theories with extended supersymmetry are most
naturally described within the harmonic superspace approach \cite{HSS,Book}.
Hence, we will consider the $\cN{=}(1,0)$ non-anti\-com\-mu\-ta\-tive charged
hypermultiplet in the harmonic superspace that was studied on the classical
level in \cite{ILZ,IZ}.
We are interested in the non-anticommutative deformation of the
holomorphic effective action that was discussed in \cite{Buch,Eremin}
using the harmonic superspace approach.
Here we generalize the results of these works to the
non-anticommutatively deformed hypermultiplet theory and compute the
leading contributions to the effective action.

One of our main results is the proof of one-loop
renormalizability of the deformed charged hypermultiplet system.
It supports the idea that the non-anticommutative deformations in general
do not spoil the renormalizability of supersymmetric theories.
Next, we find the leading contributions to the (anti)holomorphic effective
action including non-anticommutative corrections. We observe that
the holomorphic and antiholomorphic pieces are deformed differently:
the holomorphic piece is nothing but the star-generalization of
the standard holomorphic potential while the antiholomorphic piece is not.
We study also the component structure of the deformed effective action
and derive the corrections to the standard terms in the (anti)holomorphic
potential for the bosonic component fields. The deformation of the mass
and the central charge due to non-anticommutativity are found as well.

The paper is organized as follows. In Sect.~2 we review the basic
aspects of the undeformed charged hypermultiplet theory for later
comparison with the deformed ones. Sect.~3 summarizes the
most essential points about the $\cN{=}(1,0)$ non-anticommutative charged
hypermultiplet on the classical level. In Sect.~4 the leading contributions
to the low-energy effective action are computed, culminating in the
renormalizability proof. In Sect.~5 we derive the
component structure of the (anti)holomorphic effective action found in
the previous section and analyze the new terms which appear due to the
deformation. Sect.~6 contains our comments on the non-anticommutative
deformation of mass and central charge in the charged hypermultiplet model.
In the Conclusions we discuss the obtained results and point out some tempting
unsolved problems. An Appendix collects some properties of a special function
which encodes the deformation of the holomorphic potential.

\vfill\eject

\setcounter{equation}0
\section{The undeformed theory}
In this section we review briefly the known results concerning the
model of charged hypermultiplet and its effective action.

The $\cN=2$ supersymmetric models are most naturally described
within the $\cN=2$ harmonic superspace approach \cite{HSS,Book}. In
particular, the classical action of charged hypermultiplet
interacting with the Abelian vector superfield is given by
\be
S=\int d\zeta^{(-4)}du\, \breve q^+(D^{++}+V^{++})q^+.
\label{e1}
\ee
Here $q^+$ and its conjugate $\breve q^+$ are complex analytic
superfields which describe the hypermultiplet, while $V^{++}$ is a
real analytic superfield which corresponds to the vector multiplet.
The integration in (\ref{e1}) is performed over $\cN=2$ analytic
superspace with the measure $d\zeta^{(-4)}du$.
The action (\ref{e1}) is invariant under the following (Abelian) gauge
transformations of superfields
\be
\delta q^+=\lambda q^+,\qquad \delta\breve q^+=-\breve
q^+\lambda,\qquad
\delta V^{++}=-D^{++}\lambda,
\label{e2}
\ee
where $\lambda$ is a real analytic superfield.

The general structure of the low-energy effective action in the
model (\ref{e1}) is given by (see, e.g., \cite{Buch,Eremin})
\be
\Gamma=\int d^4x d^4\theta\, {\cal F}(W)+\int d^4x d^4\bar\theta\,
\bar{\cal F}(\bar W)+
\int d^4x d^8\theta\, {\cal H}(W,\bar W),
\label{e3}
\ee
where ${\cal F}$ is holomorphic potential, $\bar{\cal F}$ is
antiholomorphic potential,  ${\cal H}$ is non-ho\-lo\-mor\-phic potential.
The strength superfields $W$, $\bar W$ are gauge invariant objects which
are expressed through the gauge prepotential as follows
\be
\bar W=-\frac 14 D^{+\alpha}D^+_\alpha V^{--},\qquad
W=-\frac 14 \bar D^+_{\dot\alpha}\bar D^{+\dot\alpha}V^{--}.
\label{e4}
\ee
where
\be
V^{--}(z,u)=\int du_1 \frac{V^{++}(z,u_1)}{(u^+u^+_1)^2}
\label{e6}
\ee
is a solution of the zero-curvature equation
\be
D^{++}V^{--}-D^{--}V^{++}=0.
\label{e5}
\ee

Note that in the Abelian case the superfields $W$ and $\bar W$ are
chiral and antichiral respectively
\be
D^\pm_\alpha \bar W=0,\qquad \bar D^{\pm}_{\dot\alpha}W=0.
\label{e7}
\ee
Therefore the functions ${\cal F}(W)$ and $\bar{\cal F}(\bar W)$, which are
related to each other by complex conjugation, are integrated over the chiral and
antichiral superspaces respectively.

The perturbative low-energy effective action in the model (\ref{e1}) is studied
now in details, (see, e.g., \cite{Buch}-\cite{Dragon}) where both holomorphic and
non-holomorphic contributions have been found. In particular, the
holomorphic part of the effective action which is leading in
the low-energy approximation is given by
\be
{\cal F}(W)=-\frac{1}{32\pi^2}W^2\ln \frac W\Lambda,
\label{e8}
\ee
where $\Lambda$ is some scale.

The strength superfields $W$, $\bar W$ have the following component
structure in the bosonic sector
\be
W=\phi+(\theta^+\sigma_{mn}\theta^-)F_{mn}+\ldots,
\qquad
\bar W= \bar\phi+(\bar\theta^+\tilde
\sigma_{mn}\bar\theta^-)F_{mn}+\ldots,
\label{e9}
\ee
where $\phi,\ \bar\phi$ are scalar fields, $F_{mn}=\partial_m A_n-\partial_n A_m$
is the Maxwell strength and dots stand for the terms with spinors
$\Psi^i_\alpha,\ \bar\Psi_{i\dot\alpha}$ and
auxiliary fields ${\cal D}^{kl}$ as well as the terms with spatial derivatives. The
component structure of holomorphic effective action can be most
easily derived in the following approximation
\be
\begin{array}{c}
\phi=const,\quad \bar\phi=const,\quad F_{mn}=const,\\
\Psi^i_\alpha=\bar\Psi_{i\dot\alpha}={\cal D}^{kl}=0.
\end{array}
\label{e10}
\ee
Substituting the strength superfields
(\ref{e9}) into (\ref{e3}), one gets the component structure of
the holomorphic part of the effective action
\be
\Gamma_{hol}=\int d^4x d^4\theta\, {\cal F}(W)=
-\frac1{32\pi^2}\int d^4x (F_{mn}F_{mn}+F_{mn}\tilde F_{mn})\left(
\ln\frac\phi\Lambda +\frac32
\right)+\ldots,
\label{e11}
\ee
where $\tilde F_{mn}=\frac12\varepsilon_{mnrs}F_{rs} $ and dots
stand for the higher terms which are not essential in the
approximation (\ref{e10}). The constant 3/2 in (\ref{e11}) can be
removed by the shift of the parameter $\Lambda$, however it will
be important when we will consider the deformation of (anti)holomorphic
effective action due to non-anticommutativity.
The antiholomorphic part of effective action is given by the
complex conjugation of the action (\ref{e11}).

\setcounter{equation}0
\section{Non-anticommutative charged hypermultiplet model}
The chiral singlet deformation of $\cN=(1,1)$ superspace was
introduced in \cite{singlet} and the corresponding field models were studied
in \cite{FILSZ}-\cite{my}. Such a deformation is effectively taken into account by the
star product operator
\be
\star=\exp\left[
-I\varepsilon^{\alpha\beta}\varepsilon_{ij}
\overleftarrow{Q}^i_\alpha \overrightarrow{Q}^j_\beta
\right],
\label{e12}
\ee
which should be placed everywhere instead of usual product of superfields in
the classical actions. The constant $I$ here is a parameter of
non-anticommutativity, $Q^i_\alpha$ are the supercharges. In
particular, the non-anticommutative generalization of the action
(\ref{e1}) is given by \cite{ILZ}
\be
S=\int d\zeta^{(-4)} du\, \breve q^+\star\nabla^{++}\star
q^+,
\label{e13}
\ee
where we use the notations
\be
\nabla^{++}=D^{++}+V^{++},\qquad \nabla^{--}=D^{--}+V^{--}.
\label{13.1}
\ee
This action is invariant under the following gauge transformations
\be
\delta \breve q^+=-\breve q^+\star\lambda,\qquad
\delta q^+=\lambda\star q^+,\qquad
\delta V^{++}=-D^{++}\lambda-[V^{++},\lambda]_\star.
\label{e14}
\ee
which are the non-anticommutative generalizations of the usual ones
(\ref{e2}).

In the deformed case the strength superfields $W$, $\bar W$ are
defined by the standard equations (\ref{e4}), but the superfield
$V^{--}$ is now given by a series
\be
V^{--}(z,u)=\sum_{n=1}^\infty(-1)^n \int du_1\ldots du_n
\frac{V^{++}(z,u_1)\star V^{++}(z,u_2)\star\ldots\star V^{++}(z,u_n)}{
(u^+u^+_1)(u^+_1u^+_2)\ldots(u^+_n u^+)},
\label{e15}
\ee
which solves the star-deformed zero-curvature equation \cite{FILSZ}
\be
D^{++}V^{--}-D^{--}V^{++}+[V^{++},V^{--}]_\star=0.
\label{e16}
\ee

Let us introduce the ``bridge'' superfield $\Omega(z,u)$ as a
general $\cN=(1,1)$ superfield which relates the covariant harmonic
derivatives $\nabla^{\pm\pm}$ with the plain ones $D^{\pm\pm}$:
\be
\nabla^{++}=e^{\Omega}_\star\star
D^{++}e^{-\Omega}_\star,\qquad
\nabla^{--}=e^{\Omega}_\star\star
D^{--}e^{-\Omega}_\star.
\label{e17}
\ee
The bridge superfield was originally introduced in \cite{HSS} for the undeformed
$\cN=2$ SYM theory as an operator relating the $\cN=2$ superfields
in the $\tau$- and $\lambda$-frames. Using the bridge superfield $\Omega$
one can alternatively rewrite the equation (\ref{e15}) in the
following two equivalent forms
\be
V^{--}(z,u)=\int du'\frac{e_\star^{\Omega(z,u)}\star
e_\star^{-\Omega(z,u')}\star V^{++}(z,u')}{(u^+u'^+)^2}
=\int du'\frac{V^{++}(z,u')\star
e_\star^{\Omega(z,u')}\star
e_\star^{-\Omega(z,u)}}{(u^+u'^+)^2}.
\label{e18}
\ee
The expression (\ref{e18}) can be checked directly to satisfy the
zero-curvature condition (\ref{e16}).

It is well known \cite{Book} that the {\sl free} propagator in the hypermultiplet
model (\ref{e1}) is given by the following expression
\footnote{Note that we write here (and further) the box operator $\square$ assuming that it
is nothing but the Laplacian operator rather than a d'Alambertian one since we
deal with the Euclidian space.}
\be
G_0^{(1,1)}(1|2)=-\frac1\square(D^+_1)^4(D^+_2)^4
\frac{\delta^{12}(z_1-z_2)}{(u^+_1u^+_2)^3}
\label{e18.1}
\ee
which solves the equation
$D^{++}G_0^{(1,1)}(1|2)=\delta_A^{(3,1)}(1|2)$, where
$\delta_A^{(3,1)}(1|2)$ is the analytic delta-function.
Let us define now the {\sl full} propagator in the model (\ref{e13}) as a
distribution satisfying the equation
\be
\nabla^{++}\star G^{(1,1)}(1|2)=\delta_A^{(3,1)}(1|2).
\label{e19}
\ee
The
solution of (\ref{e19}) can formally be written as
\be
G^{(1,1)}(1|2)=-\frac1{\hat\square_\star}\star
(D^+_1)^4(D^+_2)^4\left\{
e_\star^{\Omega(1)}\star e_\star^{-\Omega(2)}
\star\frac{\delta^{12}(z_1-z_2)}{(u^+_1u^+_2)^3}
\right\},
\label{e20}
\ee
where $\hat\square_\star$ is a covariant box operator
\be
\hat\square_\star=-\frac12(D^+)^4\nabla^{--}\star\nabla^{--}.
\label{e21}
\ee
Clearly, it moves an analytic superfield to another analytic one.
The operator (\ref{e21}), acting on the analytic superfield, can be
represented in the form
\begin{eqnarray}
\hat\square_\star&=&\nabla^m\star\nabla_m
-\frac 12(\nabla^{+\alpha}\star W)\star\nabla^-_\alpha
-\frac 12(\bar\nabla^+_{\dot\alpha}\star \bar W)\star\bar\nabla^{-\dot\alpha}
+\frac 14(\nabla^{+\alpha}\star\nabla^+_\alpha\star W)\star\nabla^{--}
\nonumber\\&&
-\frac 18[\nabla^{+\alpha},\nabla^-_\alpha]_\star\star W
-\frac12\{W,\bar W \}_\star.
\label{e22}
\end{eqnarray}
Here $\nabla^\pm_\alpha=D^\pm_\alpha+V^\pm_\alpha$,
$\bar\nabla^\pm_{\dot\alpha}=\bar D^\pm_{\dot\alpha}+\bar V^\pm_{\dot\alpha}$ are
covariant spinor derivatives.
Note that the expression (\ref{e22}) has a similar form as in the
undeformed theory \cite{backgr} with the simple star-modification of
multiplication of superfields. This result is not surprising since
eq. (\ref{e22}) is derived from (\ref{e21}) only
by using the (anti)commutation relations between spinor derivatives which
have the same form as in the undeformed theory.

\setcounter{equation}0
\section{Computation of low-energy effective action}

%Before passing to direct quantum computations in the model (\ref{e13}),
%we comment briefly on the general structure of low-energy effective
%action.

\subsection{General structure of low-energy effective action}

The model (\ref{e13}) is non-anticommutative deformation of the model (\ref{e1}). One
can think naively that the effective potentials ${\cal F}$, $\bar{\cal F}$,
${\cal H}$ in the effective action (\ref{e3}) will get the analogous deformation
\be
{\cal F}(W)\longrightarrow {\cal F}_\star(W),\qquad
\bar{\cal F}(\bar W)\longrightarrow \bar{\cal F}_\star(\bar W),\qquad
{\cal H}(W,\bar W)\longrightarrow {\cal H}_\star(W,\bar W).
\label{e23}
\ee
However, we will show that this assertion is not true for the
antiholomorphic potential in the sense that no any action can be
constructed with a function $\bar {\cal F}_\star(\bar W)$
integrated over the antichiral superspace. Indeed,
 the strength superfields $W$, $\bar W$ are not
(anti)chiral, but {\sl covariantly} (anti)chiral
\be
D^+_\alpha \bar W=\nabla^-_\alpha\star\bar W=0,\qquad
\bar D^+_{\dot\alpha}W=\bar\nabla^-_{\dot\alpha}\star W=0.
\label{e24}
\ee
Therefore the expression $\int d^4x d^4\bar\theta \bar{\cal F}_\star (\bar
W)$ depends on $\theta$ variables
\be
D^-_\alpha \int d^4x d^4\bar\theta\, \bar{\cal F}_\star(\bar W)
= \int d^4x d^4\bar\theta\, [\bar{\cal F}_\star(\bar
W),V^-_{\alpha}]_\star \ne0.
\label{e25}
\ee
The rhs of (\ref{e25}) does not vanish since the star-product is
not cyclic under $d^4xd^4\bar\theta$ integration. Moreover, the
expression $\int d^4x d^4\bar\theta \bar{\cal F}_\star(\bar W)$ violates
the gauge invariance. Indeed, the strength superfields transform
covariantly under the gauge transformations (\ref{e14}) of gauge
superfield
\be
\delta W=[\lambda, W]_\star ,\qquad
\delta \bar W= [\lambda, \bar W]_\star.
\label{e26}
\ee
Since the superfields $\lambda$ and $\bar W$ are not antichiral, we
have
\be
\delta \int d^4x d^4\bar\theta\, \bar{\cal F}_\star(\bar W)=
\int d^4x d^4\bar\theta\, [\lambda, \bar{\cal F}_\star(\bar
W)]_\star\ne 0.
\label{e27}
\ee
The similar remarks on the gauge invariance in holomorphic
and antiholomorphic parts of classical action in $\cN=(1/2,0)$ gauge theory
are given in \cite{Penati}.
Note also that there is no such a problem with the holomorphic potential ${\cal
F}_\star(W)$, as it is pointed out in \cite{FILSZ} for the case of
classical $\cN=(1,0)$ SYM action.

Taking into account these remarks we propose that the general
form of the low-energy effective action in the hypermultiplet model
(\ref{e13}) is given by
\be
\Gamma=\int d^4x d^4\theta\, {\cal F}_\star(W)+
\int d^4xd^8\theta\, {\cal H}_\star(V^{++},V^{--},
W,\bar W).
\label{e28}
\ee
Here we assume that the possible terms in the low-energy effective
action which correspond to the antiholomorphic potential in the
limit $I\to 0$ is included into the function ${\cal
H}_\star(V^{++},V^{--}, W,\bar W)$ integrated in full superspace.
The direct computations will specify these functions ${\cal
F}_\star$ and ${\cal H}_\star$.

For the further considerations it will be more convenient to study
the variation of effective action $\delta\Gamma$ rather then
$\Gamma$ itself. In particular, given the holomorphic part of the
action (\ref{e28})
\be
\Gamma_{hol}=\int d^4x d^4\theta\, {\cal F}_\star(W),
\label{e29}
\ee
using the same steps as in the undeformed non-Abelian $\cN=2$ supergauge theory
\cite{Dragon}, one can write its variation either in the analytic
superspace
\be
\delta\Gamma_{hol}=\int d\zeta^{(-4)} du\, \delta V^{++}\star[-\frac 14
D^{+\alpha}D^+_\alpha{\cal F}'_\star(W)],
\label{e30}
\ee
or in the full superspace
\be
\delta\Gamma_{hol}=\int d^{12}z du\, \delta V^{++}\star
V^{--}\star\frac1W\star
{\cal F}'_\star(W).
\label{e31}
\ee

\subsection{One-loop effective action}

The one-loop effective action in the model (\ref{e13}) is
defined by the following formal expression
\footnote{Note that the one-loop effective action
in the Euclidean space is given by $\Gamma=\Tr\ln S^{(2)}_{\phi\bar\phi}$ rather
then the Minkowski space expression $\Gamma= i\Tr\ln S^{(2)}_{\phi\bar\phi}\,$.
Here $S^{(2)}_{\phi\bar\phi}$ is the second mixed functional derivative of
a classical action $S[\phi,\bar\phi]$.}
\be
\Gamma=\Tr\ln\frac{\delta^2S}{\delta\breve q^+(1)\delta q^+(2)}
=\Tr\ln(\nabla^{++}\star)=-\Tr\ln\,G^{(1,1)}(1|2),
\label{e32}
\ee
where $G^{(1,1)}(1|2)$ is given by eq. (\ref{e20}).
It is easy to find the variation of (\ref{e32})
\be
\delta\Gamma=\Tr[
\delta V^{++}\star G^{(1,1)}
]=\int d\zeta^{(-4)} du\, \delta V^{++}(1)\star
G^{(1,1)}(1|2)|_{(1)=(2)}.
\label{e33}
\ee
There is an important relation between full and free hypermultiplet
propagators
\be
G^{(1,1)}(1|3)=G_0^{(1,1)}(1|3)-\int d\zeta_2^{(-4)} du_2\,
G_0^{(1,1)}(1|2)\star V^{++}(2)\star G^{(1,1)}(2|3)
\label{e34}
\ee
which can be checked directly to satisfy (\ref{e19}).
Substituting (\ref{e34}) into (\ref{e33}), we find
\be
\delta\Gamma=-\int d\zeta_1^{(-4)} du_1 d\zeta_2^{(-4)} du_2\,
\delta V^{++}(1)\star G_0^{(1,1)}(1|2)\star V^{++}(2)\star
G^{(1,1)}(2|1).
\label{e35}
\ee
Taking into account the exact form of the propagators
(\ref{e18.1},\ref{e20}), we rewrite eq. (\ref{e35}) as follows
\begin{eqnarray}
\delta\Gamma&=&-\int d\zeta_1^{(-4)} d\zeta_2^{(-4)} du_1 du_2\,
\delta V^{++}(1)\star\frac1\square(D^+_1)^4(D^+_2)^4
\frac{\delta^{12}(z_1-z_2)}{(u^+_1u^+_2)^3}
\nonumber\\&&\times
V^{++}(2)\star\frac1{\hat\square_{\star(2)}}\star
(D^+_1)^4(D^+_2)^4\left\{
e^{\Omega(2)}_\star\star e^{-\Omega(1)}_\star
\frac{\delta^{12}(z_2-z_1)}{(u^+_2u^+_1)^3}
\right\}.
\label{e36}
\end{eqnarray}
Now we take off the derivatives $(D^+_1)^4(D^+_2)^4$ from the first
delta-function to restore the full superspace measure
\begin{eqnarray}
\delta\Gamma&=&-\int d^{12}z_1 d^{12}z_2 du_1 du_2\,
\delta V^{++}(1)\star\frac1\square
\frac{\delta^{12}(z_1-z_2)}{(u^+_1u^+_2)^3}
\nonumber\\&&\times
V^{++}(2)\star\frac1{\hat\square_{\star(2)}}\star
(D^+_1)^4(D^+_2)^4\left\{
e^{\Omega(2)}_\star\star e^{-\Omega(1)}_\star\star
\frac{\delta^{12}(z_2-z_1)}{(u^+_2u^+_1)^3}
\right\}.
\label{e37}
\end{eqnarray}
The equation (\ref{e37}) is a starting point for further
calculations of different contributions to the effective action.

\subsection{Divergent part of effective action}

To derive the divergent part of the effective action
it is sufficient to consider the approximation
\be
\frac1{\hat\square_\star}\approx \frac1\square
\label{eq77}
\ee
since all other terms in the operator $\hat\square_\star$
result to higher powers of momenta in the denominator. Upon the
condition (\ref{eq77}), the variation of effective action
(\ref{e37}) simplifies essentially
\begin{eqnarray}
\delta\Gamma_{div}&=&\int d^{12}z_1 d^{12}z_2 \frac{du_1 du_2}{(u^+_1u^+_2)^6}
\delta V^{++}(1)\star\frac1\square
\delta^{12}(z_1-z_2)
\nonumber\\&&\times
V^{++}(2)\star\frac1\square
(D^+_1)^4(D^+_2)^4\left\{
e^{\Omega(2)}_\star\star e^{-\Omega(1)}_\star\star
\delta^{12}(z_2-z_1)\right\}.
\label{eq78}
\end{eqnarray}
We have to apply the identity
\be
\delta^8(\theta_1-\theta_2)(D_1^+)^4(D^+_2)^4\delta^{12}(z_1-z_2)=
(u^+_1u^+_2)^4\delta^{12}(z_1-z_2)
\label{eq79}
\ee
to shrink the integration over the Grassmann variables to a point.
Note that all the derivatives $D^+$ in eq. (\ref{eq78}) must hit the
delta function, otherwise the result is zero since there are exactly
eight such derivatives to apply (\ref{eq79}). Moreover, the presence
of star-product can not modify the relation (\ref{eq79}).
Calculating the divergent momentum integral
\be
\left[
\int d^4k\frac{1}{k^2(p+k)^2}\right]_{\rm div}
=\frac{\pi^2}{\varepsilon}
,\qquad (\varepsilon\to0)
\label{eq21}
\ee
and applying the equation (\ref{eq79}), the
integration over $d^{12}z_2$ in (\ref{eq78}) can be performed
resulting to
\be
\delta\Gamma_{div}=\frac1{16\pi^2\varepsilon}\int d^{12}z du_1\,
\delta V^{++}(z,u_1)\star\int du_2\frac{V^{++}(z,u_2)\star
e^{\Omega(z,u_2)}_\star\star
e^{-\Omega(z,u_1)}_\star}{(u^+_1u^+_2)^2}.
\label{eq80}
\ee
Using the relation (\ref{e18}) we obtain finally
\be
\delta\Gamma_{div}=\frac1{16\pi^2\varepsilon}
\int d^{12}z du\, \delta V^{++}\star V^{--}.
\label{eq81}
\ee
The variation (\ref{eq81}) can be easily integrated with the help of
eq. (\ref{e31})
\be
\Gamma_{div}=\frac1{32\pi^2\varepsilon}\int d^4x d^4\theta\, W^2.
\label{eq81_}
\ee
We see that the divergent part of effective action is proportional
to the classical action in $\cN=(1,0)$ SYM model. In this sense the
non-anticommutative charged hypermultiplet model (\ref{e13}) is
renormalizable.

\subsection{Finite part of the effective action}

Now we will derive the finite part of the effective
action of deformed hypermultiplet model. We start with the
expression (\ref{e37}) applying the following approximation
\be
\frac1{\hat\square_\star}\approx\frac1{\square-\frac12\{W,\bar W
\}_\star}.
\label{eq82}
\ee
This means that we neglect all spatial and spinor covariant
derivatives of strength superfields in the decomposition
(\ref{e22})
\be
\partial_m W=\partial_m\bar W=0,\quad
\nabla^{+\alpha}\star W=0,\quad
\bar\nabla^+_{\dot\alpha}\star\bar W=0.
\label{eq83}
\ee
Exactly such an approximation (\ref{eq83}) is sufficient for
deriving the (anti)holomorphic contributions.
Therefore the effective action (\ref{e37}) can be rewritten as
follows
\begin{eqnarray}
\delta\Gamma&=&\int d^{12}z_1 d^{12}z_2 \frac{du_1 du_2}{(u^+_1u^+_2)^6}
\delta V^{++}(1)\star\frac1\square
\delta^{12}(z_1-z_2)
\nonumber\\&&\times
V^{++}(2)\star\frac1{\square-\frac12\{W,\bar W \}_\star}
(D^+_1)^4(D^+_2)^4\left\{
e^{\Omega(2)}_\star\star e^{-\Omega(1)}_\star\star
\delta^{12}(z_2-z_1)\right\}.
\label{eq84}
\end{eqnarray}
Once again, all the derivatives $D^+$ in the second line of
(\ref{eq84}) must hit the delta-function. Applying the identity
(\ref{eq79}) the integration over $d^8\theta_2$ is performed
\begin{eqnarray}
\delta\Gamma&=&\int d^{12}z_1 d^4x_2 \frac{du_1 du_2}{(u^+_1u^+_2)^2}
\delta V^{++}(x_1,\theta,u_1)\star V^{++}(x_2,\theta,u_2)
\star e^{\Omega(2)}_\star\star e^{-\Omega(1)}_\star
\nonumber\\&&
\star\frac1\square
\delta^4(x_1-x_2)\frac1{\square-\frac12\{W,\bar W \}_\star}
\delta^4(x_2-x_1).
\label{eq85}
\end{eqnarray}
In the momentum space the second line of (\ref{eq85}) reads
\be
\int \frac{d^4k}{(2\pi)^4}\frac1{k^2}\frac1{k^2+\frac12\{W,\bar W\}_\star}
=-\frac{1}{16\pi^2}\ln_\star\left[\frac{\{W,\bar W\}_\star}{2\Lambda^2}
\right]+({\rm divergent\ term}),
\label{eq86}
\ee
where $\Lambda$ is an arbitrary constant of dimension $+1$.
Note that the integral (\ref{eq86}) has the logarithmic divergence.
We consider here only its finite part since the divergent
contribution have been calculated above. As a result, the finite
part of (\ref{eq85}) is given by
\begin{eqnarray}
\delta\Gamma&=&-\frac1{16\pi^2}\int d^{12}z du_1\,
\delta V^{++}(x,\theta,u_1)\star \int du_2
\frac{V^{++}(x,\theta,u_2)\star
e^{\Omega(x,\theta,u_2)}_\star\star
e^{-\Omega(x,\theta,u_1)}_\star}{(u^+_1u^+_2)^2}
\nonumber\\&&
\star\ln_\star\frac{\{W,\bar W\}_\star}{2\Lambda^2}.
\label{eq87}
\end{eqnarray}
Applying the identity (\ref{e18}), we conclude
\be
\delta\Gamma=-\frac1{16\pi^2}\int d^{12}z du\,
\delta V^{++}\star  V^{--}\star \ln_\star\frac{\{W,\bar W\}_\star}{
2\Lambda^2}.
\label{eq88}
\ee

The variation of the effective action (\ref{eq88}) is one of the
main results of the present work. It gives us the part of the
low-energy effective action depending on the strength superfields
without derivatives.

If the parameter of non-anticommutativity tends to zero, $I\to0$,
the equation (\ref{eq88}) reduces to the holomorphic and
antiholomorphic parts of the effective action
\be
\delta\Gamma_{(I=0)}=-\frac1{16\pi^2}\int d^{12}z du\,
\delta V^{++} V^{--} \left(\ln\frac W\Lambda +\ln \frac{\bar W}\Lambda\right).
\label{eq89}
\ee
The variation (\ref{eq89}) exactly corresponds to the holomorphic
potential (\ref{e8}) and its conjugate.

Note that when $I\ne0$, the logarithm in (\ref{eq88}) can not be
represented in a form of a sum of holomorphic and antiholomorphic
parts. Therefore, the variation of the effective action (\ref{eq88})
is responsible for both holomorphic, antiholomorphic and
non-holomorphic contributions to the effective action.

\subsection{Holomorphic contribution}

Let us single out purely holomorphic part from the expression
(\ref{eq88}). For this purpose we restrict the background strength
superfield $\bar W$ to be constant
\be
\bar W=\bar {\bf W}=const,
\label{eq90}
\ee
and the log function in (\ref{eq88}) simplifies to
\be
\ln_\star\frac{\{W,\bar W \}_\star}{2\Lambda^2}=
\ln_\star\frac{W}\Lambda+\ln\frac{\bar{\bf W}}\Lambda.
\label{eq91}
\ee
The holomorphic part is now given by
\be
\delta\Gamma_{hol}=-\frac1{16\pi^2}\int d^{12}z du\,
\delta V^{++}\star  V^{--}\star \ln_\star\frac W\Lambda.
\label{eq93.1}
\ee
According to the equation (\ref{e31}), the variation
(\ref{eq93.1}) can easily be integrated:
\be
\Gamma_{hol}=-\frac1{32\pi^2}\int d^4x d^4\theta\, W\star W\star
\ln_\star \frac{W}{\Lambda}.
\label{eq93.2}
\ee
As a result, we proved that the holomorphic part of effective action in
the hypermultiplet model is nothing but a star-product
generalization of a standard holomorphic potential (\ref{e8}).

\subsection{Antiholomorphic contribution}

Similarly, the antiholomorphic part of the effective action can be found
from (\ref{eq88}) when we restrict the strength $W$ to be constant
\be
W={\bf W}=const.
\label{e93.2.1}
\ee
The antiholomorphic part now reads
\be
\delta\Gamma_{antihol}=\frac1{16\pi^2}\int d^{12}z du\,
\delta V^{++}\star  V^{--}\star \ln_\star\frac{\bar W}\Lambda.
\label{eq93.3}
\ee
In contrast to the variation (\ref{eq93.1}), the expression
(\ref{eq93.3}) can not be so easily integrated since there is no
antiholomorphic potential written in the antichiral superspace.
However, one can readily find the (part of) effective equation of
motion corresponding to the variation (\ref{eq93.3})
\be
\frac{\delta \Gamma_{antihol}}{\delta V^{++}}
=\frac1{16\pi^2}(\bar D^+)^2\left[ \bar W\star \ln_\star\frac{\bar
W}\Lambda
\right].
\label{eq93.4}
\ee
The variation (\ref{eq93.3}) will be integrated in the next section only
for some particular choice of background gauge superfields.

\setcounter{equation}0
\section{Component structure of low-energy effective action}
We are interested in the leading component terms of the
(anti)holomorphic effective actions (\ref{eq93.2}) and (\ref{eq93.3})
in the bosonic sector. Clearly, these effective actions should give
the corrections to the corresponding expression (\ref{e11}) in the
undeformed theory. Therefore we will use the same ansatz
(\ref{e10}) for the component fields. Here we will follow the works
\cite{FILSZ,ILZ}, using the same conventions and notations for the
component fields and sigma-matrices.

The scalar and vector fields enter the prepotential $V^{++}$ in the
Wess-Zumino gauge as follows
\be
V^{++}_{WZ}=(\theta^+)^2\bar\phi+(\bar\theta^+)\phi
+(\theta^+\sigma_m\bar\theta^+)A_m
-2i(\bar\theta^+)^2(\theta^+\theta^-)\partial_m A_m
-(\bar\theta^+)^2(\theta^-\sigma_{mn}\theta^+)F_{mn}.
\label{eqq100}
\ee
The prepotential $V^{--}$ is defined as a solution
of zero-curvature equation (\ref{e16}).
Unfolding the star-product in (\ref{e16}), we have
\begin{eqnarray}
&&D^{++}V^{--}-D^{--}V^{++}_{WZ}+2I[\partial_+^\alpha V^{++}_{WZ}\partial_{-\alpha}V^{--}
-\partial_-^\alpha V^{++}_{WZ}\partial_{+\alpha}V^{--}]
\nonumber\\
&&+\frac{I^3}2[\partial_-^\alpha(\partial_+)^2V^{++}_{WZ}
\partial_{+\alpha}(\partial_-)^2V^{--}
-\partial_+^\alpha(\partial_-)^2V^{++}_{WZ}\partial_{-\alpha}(\partial_+)^2V^{--}]=0,
\label{eqq102}
\end{eqnarray}
where
\be
\partial_{+\alpha}=\frac\partial{\partial\theta^{+\alpha}},\qquad
\partial_{-\alpha}=\frac\partial{\partial\theta^{-\alpha}}.
\label{eqq103}
\ee
One can look for the prepotential $V^{--}$ in the following form
\begin{eqnarray}
V^{--}&=&v^{--}+\bar\theta^-_{\dot\alpha}v^{-\dot\alpha}+
(\bar\theta^-)^2 A
+(\bar\theta^+\bar\theta^-)\varphi^{--}\nonumber\\
&&+(\bar\theta^+\tilde\sigma^{mn}\bar\theta^-)\varphi^{--}_{mn}
+(\bar\theta^-)^2\bar\theta^+_{\dot\alpha}\tau^{-\dot\alpha}
+(\bar\theta^+)^2(\bar\theta^-)^2\tau^{--},
\label{eqq107}
\end{eqnarray}
where all fields in the rhs of eq. (\ref{eqq107}) depend only on
$\theta^+_\alpha,\ \theta^-_\alpha$ variables. The superfields
$v^{--}$, $v^{-\dot\alpha}$, $\varphi^{--}$, $A$,
$\tau^{-\dot\alpha}$, $\tau^{--}$ should be found from the eq.
(\ref{eqq102}).
The iterative procedure of solving eq. (\ref{eqq102}) is given in
\cite{FILSZ}. Following the same steps we find
\begin{eqnarray}
v^{--}&=&(\theta^-)^2\frac{\bar\phi}{1+4I\bar\phi},
\label{eqq108}\\
v^{-\dot\alpha}&=&\frac{(\theta^-\sigma_m)^{\dot\alpha}A_m}{1+4I\bar\phi},
\label{eqq109}\\
\varphi^{--}&=&-\frac{2i(\theta^-)^2\partial_mA_m}{1+4I\bar\phi},
\label{eqq110}\\
A&=&\phi+\frac{4IA_mA_m}{1+4I\bar\phi}
+(\theta^+\sigma_{mn}\theta^-)F_{mn},
\label{eqq111}\\
\tau^{-\dot\alpha}&=&\frac{4I(\theta^-\sigma_{mn})^\alpha F_{mn}\sigma_{
r\alpha}{}^{\dot\alpha}A_r}{1+4I\bar\phi},
\label{eqq112}\\
\tau^{--}&=&\frac{4I(\theta^-)^2(F_{mn}F_{mn}+F_{mn}\tilde
F_{mn})}{1+4I\bar\phi}.
\label{eqq113}
\end{eqnarray}

Now we obtain the component structure of the strength superfields
\begin{eqnarray}
W&=&-\frac14(\bar D^+)^2V^{--}=\phi+\frac{4IA_m
A_m}{1+4I\bar\phi}+(\theta^+\sigma_{mn}\theta^-)F_{mn},
\label{eqq114}\\
\bar W&=&-\frac14( D^+)^2V^{--}=\frac{\bar\phi}{1+4I\bar\phi}+
(\bar\theta^+\tilde\sigma_{mn}\bar\theta^-)\frac{F_{mn}}{1+4I\bar\phi}.
\label{eqq115}
\end{eqnarray}
Note that the strength superfields (\ref{eqq114}) and (\ref{eqq115}) are
deformed differently. It is clear that in the limit $I\to0$ these
expressions coincide with the undeformed ones
(\ref{e9}). Introducing the notations
\be
\Phi=\phi+\frac{4IA_m A_m}{1+4I\bar\phi},\qquad
\bar\Phi=\frac{\bar\phi}{1+4I\bar\phi},\qquad
{\bf F}_{mn}=\frac{F_{mn}}{1+4I\bar\phi},
\label{eqq116}
\ee
the eqs. (\ref{eqq114},\ref{eqq115}) can be written in a form
similar to eq. (\ref{e9})
\be
W=\Phi+(\theta^+\sigma_{mn}\theta^-)F_{mn},\qquad
\bar W=\bar\Phi+(\bar\theta^+\tilde\sigma_{mn}\bar\theta^-){\bf
F}_{mn}.
\label{eqq117}
\ee

To find the component structure of the holomorphic potential
(\ref{eq93.2},\ref{eq93.3}) we have to compute the following
quontities
\begin{eqnarray}
W\star W&=&\Phi^2+4I^2(F^2+F\tilde F)+
2(\theta^+\sigma_{mn}\theta^-)F_{mn}\Phi
+(\theta^+)^2(\theta^-)^2(F^2+F\tilde F),
\label{eqq121}\\
\ln_\star\frac W\Lambda&=&\ln\frac\Phi\Lambda+\frac14\ln\left[
1-\frac{8I^2(F^2+F\tilde F)}{\Phi^2}\right]
+(\theta^+\sigma_{mn}\theta^-)\frac{F_{mn}}{\Phi}
\frac{{\rm arcth}\sqrt{\frac{8I^2(F^2+F\tilde F)}{\Phi^2}}}{
\sqrt{\frac{8I^2(F^2+F\tilde F)}{\Phi^2}}}
\nonumber\\&&
+\frac1{16I^2}(\theta^+)^2(\theta^-)^2\ln\left[
1-\frac{8I^2(F^2+F\tilde F)}{\Phi^2}
 \right],
\label{eqq122}\\
\bar W\star\bar W&=&\bar W\bar W=
\bar\Phi^2+
2(\bar\theta^+\tilde\sigma_{mn}\bar\theta^-)F_{mn}\bar\Phi
+(\bar\theta^+)^2(\bar\theta^-)^2(F^2+F\tilde F),
\label{eqq122.1}\\
\ln_\star\frac{\bar W}\Lambda&=&
\ln\frac{\bar W}{\Lambda}=
\ln\frac{\bar\Phi}\Lambda
+(\bar\theta^+\tilde\sigma_{mn}\bar\theta^-)
\frac{{\bf F}_{mn}}{\bar\Phi}
-\frac12(\bar\theta^+)^2(\bar\theta^-)^2\frac{{\bf F}^2+{\bf F}\tilde {\bf
F}}{\bar\Phi^2}.
\label{eqq122.2}
\end{eqnarray}
As a result, substituting the expressions
(\ref{eqq121},\ref{eqq122}) into eq. (\ref{eq93.2}),
we find the component structure of the holomorphic effective action
\be
\Gamma_{hol}=-\frac1{32\pi^2}\int d^4x (F^2+F\tilde F)
\left[\ln\frac{\Phi}{\Lambda}+\Delta(X(\Phi,F_{mn})) \right],
\label{eqq124}
\ee
where
\begin{eqnarray}
\Delta(X)&=&\frac12(1-X)^2
\ln(X-1)+\frac12(1+ X)^2\ln(1+ X)-(1+X^2)\ln X,
\label{eqq126}\\
X(\Phi,F_{mn})&=&\frac\Phi{2I\sqrt{2(F^2+F\tilde
F)}}.
\label{eqq124.1}
\end{eqnarray}
The equation (\ref{eqq124}) shows that the function
$\Delta(X)$ is a correction due to non-an\-ti\-com\-mu\-ta\-ti\-vi\-ty
to the standard holomorphic effective action (\ref{e11}).
This function has the smooth limit at $I\to0$
\be
\lim_{I\to0}\Delta(X)=\frac32.
\label{eqq127}
\ee
Note that exactly the constant 3/2 stands in the rhs in eq.
(\ref{e11}) which was not essential in the undeformed theory, but
now this constant is replaced by the function $\Delta(X)$.
More detailed studies of the function $\Delta(X)$ are given in the
Appendix.

Let us consider the antiholomorphic potential when the strength
superfield $\bar W$ is defined by the component expression
(\ref{eqq115}) with the fields $F_{mn}$ and $\bar\phi$ being
constant. The equations (\ref{eqq122.1},\ref{eqq122.2}) show that the
star-product can be omitted for such an approximation. Therefore the
variation (\ref{eq93.3}) can be integrated is the same way as in the
undeformed theory
\begin{eqnarray}
\Gamma_{antihol}&=&-\frac1{32\pi^2}\int d^4x d^4\bar\theta
\,\bar W^2\ln\frac{\bar W}{\Lambda}
=-\frac1{32\pi^2}\int d^4x ({\bf F}^2+{\bf F}\tilde {\bf F})
 \left(\ln\frac{\bar\Phi}{\Lambda}+\frac32\right)
 \nonumber\\&&
=-\frac1{32\pi^2}\int d^4x \frac{(F^2+F\tilde F)}{
(1+4I\bar\phi)^2}\left(
 \ln\frac{\bar\phi}{\Lambda(1+4I\bar\phi)}+\frac32\right).
\label{eqq135}
\end{eqnarray}
Here the equations (\ref{eqq122.1},\ref{eqq122.2},\ref{eqq116}) have been used.
We see that the non-anticommutativity manifests itself here by a simple
rescaling of fields by the factor $1/(1+4I\bar\phi)$.
In the limit $I\to0$ the expression (\ref{eqq135}) reduces to the
standard one for the antiholomorphic potential.

\section{Deformation of the central charge and the mass}
It is well known \cite{Buch} that the model of hypermultiplet interacting
with the external vector superfield possesses non-trivial central charge which
is related to the mass of the hypermultiplet via BPS relation. Let us find the deformation of
central charge and the mass in the case of non-anticommutative singlet
deformation under considerations.

The central charges in the $\cN=2$ theories
arise effectively from non-vanishing vacuum expectation values of scalar
fields \cite{Buch}
\be
\langle\phi\rangle=a,\qquad \langle\bar\phi\rangle=\bar a.
\label{eq94_}
\ee
The constants $a$, $\bar a$ enter the prepotentials
(\ref{eqq100},\ref{eqq107}) as follows
\begin{eqnarray}
{\bf V}^{++}&=&a(\bar\theta^+)^2+\bar a(\theta^+)^2,
\label{eq94}\\
{\bf V}^{--}&=&a(\bar\theta^-)^2+\frac{\bar a}{1+4I\bar a}(\theta^-)^2.
\label{eq95}
\end{eqnarray}
The corresponding strength superfields (\ref{eqq114},\ref{eqq115})
read now
\be
{\bf W}=a,\qquad \bar{\bf W}=\frac{\bar a}{1+4I\bar a}.
\label{eq96}
\ee
We see that the strength $\bar {\bf W}$ is deformed by the non-anticommutativity
while ${\bf W}$ is not.

Now, using the standard relations $V^-_{\alpha}=-D^+_\alpha V^{--}$,
$\bar V^-_{\dot\alpha}=-\bar D^+_{\dot\alpha} V^{--}$, we derive the covariant
spinor derivatives, corresponding to the prepotentials (\ref{eq94},\ref{eq95})
\be
\begin{array}{ll}
\bar{\bf D}^+_{\dot\alpha}=\bar D^+_{\dot\alpha},
 & {\bf D}^+_\alpha=D^+_\alpha, \\
\bar{\bf D}^-_{\dot\alpha}=\bar D^-_{\dot\alpha}+2a\bar\theta^-_{\dot\alpha},
\quad &
{\bf D}^-_\alpha=D^-_\alpha-\dfrac{2\bar a}{1+4I\bar a}\theta^-_\alpha.
\end{array}
\label{eq97}
\ee
The supercharges anticommuting with the derivatives (\ref{eq97}) are
\be
\begin{array}{ll}
\bar {\bf Q}^+_{\dot\alpha}=\bar Q^+_{\dot\alpha}-2a\bar\theta^+_{\dot\alpha},
\quad &
{\bf Q}^+_\alpha=Q^+_\alpha+\dfrac{2\bar a}{1+4I\bar a}\theta^+_\alpha,\\
\bar {\bf Q}^-_{\dot\alpha}=\bar Q^-_{\dot\alpha},&
{\bf Q}^-_\alpha=Q^-_\alpha.
\end{array}
\label{eq98}
\ee
Note that only the supercharge ${\bf Q}^+_\alpha$ is deformed by the
non-anticommutativity. It is easy to find now the anticommutation relation
between the supercharges (\ref{eq98})
\be
\{\bar {\bf Q}^+_{\dot\alpha},\bar {\bf Q}^-_{\dot\beta}
\}=2\bar Z\varepsilon_{\dot\alpha\dot\beta},\qquad
\{{\bf Q}^+_\alpha,{\bf Q}^-_\beta \}=-2Z\varepsilon_{
\alpha\beta},
\label{eq99}
\ee
where the central charges $Z$, $\bar Z$ are given by
\be
\bar Z=a,\qquad Z=\frac{\bar a}{1+4I\bar a}.
\label{eq100}
\ee
As a result, only the central charge $Z$ is deformed by the
non-anticommutativity while $\bar Z$ is not.

It is well known that in the presence of central charge the hypermultiplet
acquires the BPS mass. To find the mass, let us consider the operator
\be
\hat\ssquare=-\frac12(D^+)^4(\nnabla^{--})^2=\square-m^2,
\label{eq101}
\ee
where $\nnabla^{--}=D^{--}+{\bf V}^{--}$ and
\be
m^2=Z\bar Z=\frac{a\bar a}{1+4I\bar a}.
\label{eq102}
\ee
It is easy to see that $m^2$, given by eq. (\ref{eq102}), is a mass
squared of the hypermultiplet. Indeed, let us consider the
hypermultiplet model interacting with the background vector
superfield (\ref{eq94})
\be
{\bf S}=\int d\zeta^{(-4)} du\,\breve q^+\star(D^{++}+{\bf V}^{++})\star
q^+.
\label{eq103}
\ee
The model (\ref{eq103}) is effectively described by the massive propagator
\be
{\bf G}^{(1,1)}(1|2)=-\frac1{\square-m^2}(D^+_1)^4(D^+_2)^4
\left\{e^{{\bf \Omega}(1)}_\star\star e^{-{\bf \Omega}(2)}_\star\star
\frac{\delta^{12}(z_1-z_2)}{(u^+_1u^+_2)^3} \right\},
\label{eq104}
\ee
which can be derived in the same way as the general one (\ref{e20}).
Here $\bf\Omega$ is a bridge superfield corresponding to the
prepotential ${\bf V}^{++}$ (\ref{eq94}).

Note that the mass squared (\ref{eq102}) and the central charges
(\ref{eq100}) are deformed in such a way that the BPS relation $m^2=Z\bar
Z$ is conserved.

\section{Conclusions}
In this paper we studied the low-energy effective action
and renormalizability of $\cN{=}(1,0)$ non-anticommutative charged
hypermultiplet theory. This model describes the interaction of a
hypermultiplet with an Abelian vector superfield under the
singlet chiral deformation of supersymmetry. Let us summarize the
basic results obtained in the present work.
\begin{enumerate}
\item The procedure of perturbative quantum computation of the
low-energy effective action for the deformed charged hypermultiplet model
is developed within the harmonic superspace approach.
\item Using this procedure, the divergent part of the effective
action is calculated and is shown to be proportional to the
classical action of $\cN{=}(1,0)$ non-anticommutative SYM theory. In
this sense the present model is renormalizable.
\item The general structure of the low-energy effective action of this
theory is revealed. Away from the undeformed limit, the antiholomorphic piece
no longer exists by itself but is incorporated in a
{\sl full\/} $\cN{=}(1,1)$ superspace integral.
\item The holomorphic effective action is calculated and remains a
{\sl chiral\/} superspace integral. It is shown
to be given by the holomorphic potential which is a
star-generalization of the undeformed one. The contribution to
(the variation of) the effective action that
corresponds to the antiholomorphic potential is also found as an
expression written in full $\cN{=}(1,1)$ superspace.
\item The component structure of the (anti)holomorphic effective action is
studied in the bosonic sector in the constant-fields approximation.
It is shown that the holomorphic and antiholomorphic potentials still get
deformed differently. In this approximation the deformed
holomorphic potential (\ref{eqq124}) acquires the extra terms
given by the function (\ref{eqq126}). For the antiholomorphic piece, it is
shown that
the deformation merely effects a rescaling of component fields by a factor of
$(1+4I\bar\phi)^{-1}$, where $\bar\phi$ is one of the two scalar fields of
the vector multiplet.
\item The deformation of the mass and the central charge is found.
It is shown that the mass-squared as well as the central charge are rescaled
by the same factor $(1+4I\bar\phi)^{-1}$, preserving the relation between them.
\end{enumerate}

In the light of the present results, it would be rewarding to
solve the following problems concerning the quantum aspects of
non-anticommutative theories with extended supersymmetry. First, it
is tempting to determine for the hypermultiplet model the deformation
of the next-to-leading terms in the effective action, which are
necessarily non-holomorphic. Also, one should
develop the non-Abelian generalization. Next, it is important to perform
an analogous investigation for the pure SYM theory, since in the undeformed
case this model (the Seiberg-Witten theory)
plays an important role in modern theoretical physics. Finally, it
would be interesting to extend the quantum studies of
non-anticommutative theories to the case of non-singlet (i.e.~more general)
deformations of supersymmetry, as considered particularly in
\cite{CILQ} on the classical level.

\section*{Acknowledgements}
The present work is supported particularly by the
DFG grant, project No 436 RUS 113/669/0-3 and by INTAS grant.
The work of I.L.B and I.B.S. was supported by RFBR grant,
project No 06-02-16346, joint RFBR-DFG grant, project No 06-02-04012
and LRSS grant, project No 4489.2006.2.
IBS acknowledges the support from the grant of the President of
Russian Federation, project No 7110.2006.2.
O.L. acknowledges support from
the DFG grant Le-838/7.

\def\theequation{A.\arabic{equation}}
\setcounter{equation}0
\appendix
\section{Appendix}

Let us give some more comments about the function
$\Delta(X)$ given by eq. (\ref{eqq126}).

{\bf i.} Due to the presence of log function, the expression (\ref{eqq126}) is
well-defined if $X>1$, or
\be
\Phi>2I\sqrt{2(F^2+F\tilde F)}.
\label{eqq128}
\ee
The equation (\ref{eqq128}) means that the vacuum values for the
scalar field $\phi$ are bounded below if $I\ne0$.

{\bf ii.} In the region (\ref{eqq128}) the function $3/2-\Delta(X)$
is monotone decreasing. It is plotted in the Fig.
1a.

\begin{figure}[tb]
\begin{center}
\epsfxsize=15cm
\epsfbox{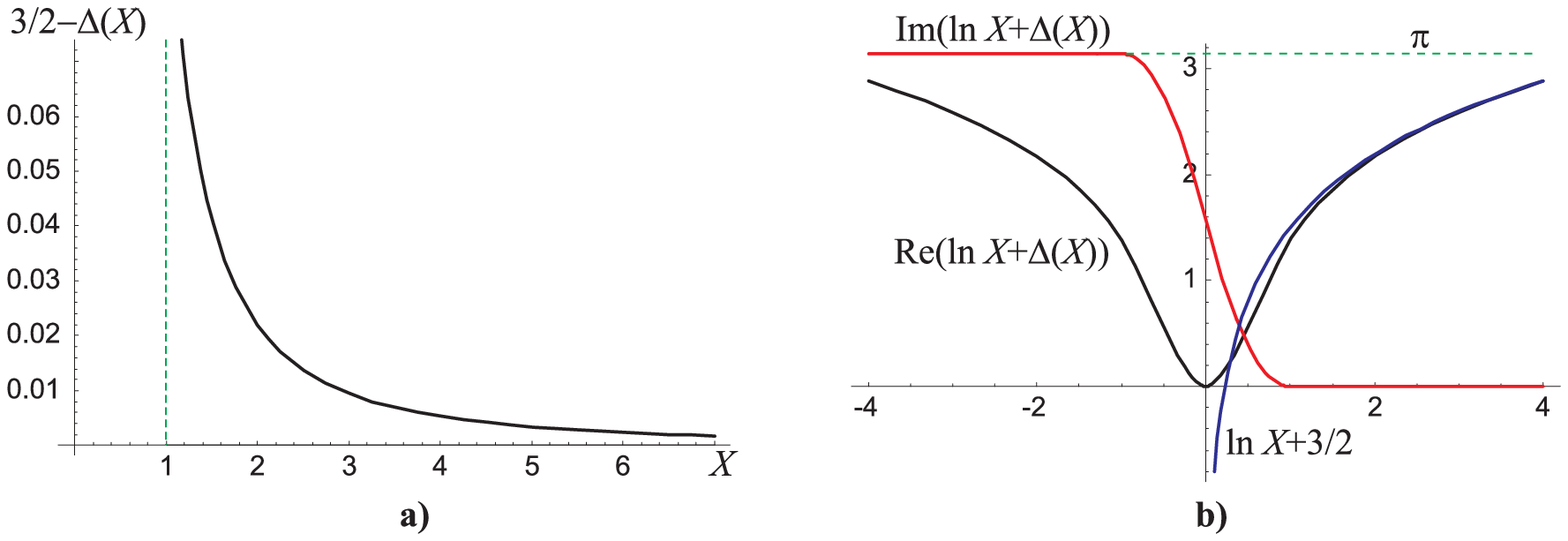}
\end{center}
\caption{}
\end{figure}

{\bf iii.} If one relaxes the condition (\ref{eqq127}), the
function $\Delta(X)$, as well as the holomorphic effective action,
acquires an imaginary part. The expressions ${\rm Re}(\ln
X+\Delta(X))$, ${\rm Im}(\ln X+\Delta(X))$ are plotted in Fig. 1b
in comparison with the function $\ln X+3/2$ which is responsible for
the bosonic part of the holomorphic effective action (\ref{e11}) in the
undeformed case. It is interesting to note that the function
$\ln X+\Delta(X)$ has no logarithmic singularity at the origin $X=0$
when $I\ne0$.

\end{document}